\documentclass[superscriptaddress,secnumarabic,twocolumn,
 amssymb,amsmath,nobibnotes,aps,prd,showkeys,showpacs,nofootinbib]{revtex4}
\usepackage{mathtools}
\usepackage{graphicx}
\usepackage{booktabs}
\usepackage{diagbox}
\usepackage{appendix}
\usepackage{slashbox}
\usepackage{color}
\usepackage[normalem]{ulem}
\usepackage[utf8]{inputenc}

\begin{document}

\title{Spinning test body orbiting around a Kerr black hole: Comparing Spin Supplementary Conditions for Circular Equatorial Orbits  }
\author{Iason Timogiannis}
\email{timogian@phys.uoa.gr}
\affiliation{Section of Astrophysics, Astronomy, and Mechanics, Department of Physics,
University of Athens, Panepistimiopolis Zografos GR15783, Athens, Greece}
\author{Georgios Lukes-Gerakopoulos}
\email{gglukes@gmail.com}
\affiliation{Astronomical Institute of the Czech Academy of Sciences,
Bo\v{c}n\'{i} II 1401/1a, CZ-141 00 Prague, Czech Republic}
\author{Theocharis A. Apostolatos}
\email{thapostol@phys.uoa.gr}
\affiliation{Section of Astrophysics, Astronomy, and Mechanics, Department of Physics,
University of Athens, Panepistimiopolis Zografos GR15783, Athens, Greece}

\begin{abstract}
The worldline of a spinning test body moving in curved spacetime can be provided by the Mathisson-Papapetrou-Dixon (MPD) equations when its centroid, i.e.\ its center of mass, is fixed by
a Spin Supplementary Condition (SSC).  In the present study, we continue the exploration of shifts between different centroids started in a recently published work [{\it Phys. Rev. D} {\bf 104}, 024042 (2021)], henceforth Paper I, for the Schwarzschild spacetime, by examining the frequencies of circular equatorial orbits under a change of the SSC in the Kerr spacetime. In particular, we examine the convergence in the terms of the prograde and retrograde orbital frequencies, when these frequencies are expanded in power series of the spin measure and the centroid of the body is shifted from the Mathisson-Pirani or the Ohashi-Kyrian-Semer\'{a}k frame to the Tulczyjew-Dixon one. Since in Paper I, we have seen that the innermost stable circular orbits (ISCOs) hold a special place in this comparison process, we focus on them rigorously in this work. We introduce a novel method of finding ISCOs for any SSC and employ it for the Tulczyjew-Dixon and the Mathisson-Pirani formalisms. We resort to numerical investigation of the convergence between the SSCs for the ISCO case, due to technical difficulties  not allowing Paper's I analytical treatment. Our conclusion, as in Paper I, is that there appears to be a convergence in the power series of the frequencies between the SSCs, which is improved when the proper shifts are taken into account, but there exists a limit in this convergence due to the fact that in the spinning body approximation we consider only the first two lower multipoles of the extended body and ignore all the higher ones. 
\end{abstract}

\pacs{~~}
\maketitle

\section{Introduction}
The two body problem is a fascinating, yet challenging problem in general relativity. Having to simultaneously determine the motion and the gravitational field of a binary system, which are governed by the field equations \cite{Einstein15,Einstein16}, pushes both analytical and numerical methods to their limits \cite{Blanchet14,Barack19,Brugmann18}. Technical and conceptual challenges arise even in the case that the presence of one of the bodies is prescribed by a fixed spacetime background, while the other is approximated as a test body. In this seemingly simple limit, one has to still find a way to describe the motion of an extended test body in a curved spacetime background. To tackle this issue, the pioneering works of Mathisson \cite{Mathisson37}, Papapetrou \cite{Papapetrou51} and Dixon \cite{Dixon64} have provided a theoretical framework, in which the extended body's structure is described by a series of multipole moments. In the simplest setup of this framework, only the first two multipoles are taken into account, i.e. the pole and the dipole, while the quadrupolar as well as higher moments are neglected. In this pole-dipole approximation the body has apart from its mass an internal angular momentum, i.e. a spin. The equations of motion of this spinning test body, if we consider only gravitational interactions, reduce the Mathisson-Papapetrou-Dixon (MPD) equations\footnote{The MPD equations can be obtained through the covariant conservation of the stress-energy tensor $T^{\mu\nu}$ of the extended body.} \cite{Dixon74b} to the following set of equations:
\begin{align}
&\frac{Dp^\mu}{d\lambda} =-\frac{1}{2}R^\mu_{\nu\kappa\lambda} u^\nu S^{\kappa \lambda}, \label{eq:MPD_mom}\\
&\frac{DS^{\mu \nu}}{d\lambda} =p^\mu u^\nu -u^\mu p^\nu, \label{eq:MPD_spin}
\end{align}
which evolve the four-momentum $p^\mu$ along with the spin tensor $S^{\mu\nu}$ of the test body and where $\frac{D}{d\lambda} := u^\mu \nabla_\mu$ denotes the covariant derivative in the direction of the four-velocity $\displaystyle u^\mu=\frac{d z^\mu (\lambda)}{d\lambda}$, since we choose $\lambda$ to be the proper time; $z^\mu(\lambda)$ is the coordinate position of the representative worldline of the test body. The rest mass of the extended spinning body can be dually defined, either with respect to the four-momentum $\mu:=\sqrt{-p^\nu p_\nu}$ (dynamical mass) or with respect to the four-velocity $m:=-p^\nu u_\nu$ (kinematical mass). The measure of its spin, by contrast, is uniquely defined as:  
\begin{align} \label{eq:spin_m}
 S^2=\frac{1}{2} S^{\mu\nu}S_{\mu\nu}\, ,
\end{align}
which discloses the spacelike character of the antisymmetric spin tensor $S^{\mu\nu}$. 

The underdetermination issue of the MPD set of equations has been thoroughly examined since the initial derivation of the equations. As a result, a great abundance of different constraints, known as spin supplementary conditions (SSCs), has been developed in order to address it (see for instance \cite{Semerak15}, which can serve as a relatively recent review on SSCs). In brief, the SSCs required to close the MPD system of equations fix the centroid of the body, whose evolution in time forms the representative worldline. The centroid serves as the point against which the spin is calculated. All the established SSCs can be written covariantly in the form $V_\mu S^{\mu\nu}=0$, where $V^\nu$ stands for the reference time-like vector, which is often normalized to $V^\nu V_\nu=-1$, like the test body's four-velocity does. Once a SSC is imposed, it is possible to introduce a spin four-vector by means of the Levi-Civita tensor $\epsilon_{\mu\nu\rho\sigma}$, that is:
\begin{align}
\label{eq:SpinVect}
 S_\mu := -\frac{1}{2} \epsilon_{\mu\nu\rho\sigma}
          \, V^\nu \, S^{\rho\sigma} \; ,
\end{align}
while the inverse relation of Eq.~\eqref{eq:SpinVect} reads:
\begin{align}
    S^{\rho\sigma}=-\epsilon^{\rho\sigma\nu\kappa}S_\nu V_\kappa.
\end{align}

The pole-dipole version of the Mathisson-Papapetrou-Dixon formalism appears to be adequate for modelling an Extreme Mass Ratio Inspiral (EMRI) \cite{Piovano20,Mathews22}, i.e.\ a compact astrophysical object, like a neutron star or a stellar mass black hole, captured into an inspiral orbit, around a central supermassive black hole. The latter are extremely massive objects residing at the core of many galaxies, including our own Milky Way. EMRIs are among the prime targets for space-based gravitational wave detectors like LISA (Laser Interferometer Space Antenna) \cite{EMRIsLISA} and as such, play a vital role in the gravitational research which will be conducted in the couple of decades to come until its launch. 

In Paper I \cite{Timogiannis21}, we probed the impact of the centroid's alteration, on the characteristic features of an extended spinning test body, expanded in power series with respect to its spin measure, within the context of EMRIs. Namely, we examined whether the orbital frequency of an inspiralling body, moving in an arbitrary circular equatorial orbit around a massive non-rotating uncharged black hole is preserved under the transition to another centroid of the same physical body governed by a different SSC. One of the main concluding remarks of Paper I was that the appropriate shift between centroids is not a gauge transformation as in flat spacetime for the pole-dipole approximation and consequently the convergence between the discussed SSCs holds only up to quadratic spin terms and cannot be achieved for the whole power series. The primary objective of the present work remains almost the same. More precisely, we extend our investigations to a more general spacetime, i.e.\ the Kerr spacetime background, by testing if the spherical symmetry breaking affects the degree of this convergence between three particular SSCs: 
\begin{itemize}
    \item The Tulczyjew-Dixon (TD) SSC \cite{Dixon70,Tulczyjew59} which uses $V^\nu:=p^\nu/\mu$ as a future oriented timelike vector and under which $\mu$, $S$ are constants of motion, independently of the background spacetime \cite{Semerak99,LG17}.  \item The Mathisson-Pirani (MP) SSC \cite{Mathisson37,Pirani56} which uses $V^\nu:=u^\nu$ as a future oriented timelike vector and under which $m$, $S$ are constants of motion, independently of the background spacetime \cite{Semerak99,LG17}.
    \item The Ohashi-Kyrian-Semer\'{a}k (OKS) SSC \cite{Ohashi03,Kyrian07,Semerak15} which promotes $V^\nu$ to an additional variable of the system and an evolution equation $\frac{D V^\nu}{d \lambda}=0$ is defined for it. Under this SSC the two different notions of mass are identical, i.e.\ $\mu=m$, and constant upon evolution along with $S$ \cite{Semerak99,Semerak15}. Moreover, under OKS SSC the test body's four-momentum and four-velocity are correlated linearly, or in other words $p^\nu=\mu u^\nu=m u^\nu$, in complete analogy to the geodesic limit\footnote{Note that for spinning bodies $p^\nu$ and $u^\nu$ are not necessarily parallel. Their relation is given by $p^\nu=\mu u^\nu+u_\mu \frac{DS^{\mu \nu}}{d\lambda}$, with the second term on the right hand side known as the hidden momentum.}.  
\end{itemize}
Besides the SSC-dependent constants of motion, there are the spacetime dependent integrals of motion originating from a Killing vector $\xi^\mu$, which expresses a background's symmetry. For an extended test body such an integral of motion reads:
\begin{equation}
C=\xi^\mu p_\mu-\frac{1}{2}\xi_{\mu;\nu}S^{\mu\nu},
\end{equation}
(see \cite{Dixon70} for a derivation). Hence, the stationarity and the axisymmetry of the Kerr metric leads to the conservation of the energy $E$ and of the z-component of the total angular momentum of the spinning body $J_z$ respectively. These quantities can be written as \cite{Harms16}: 
\begin{align}
    &E=-p_t+\frac{S}{2}\sqrt{-\frac{g_{\theta\theta}}{g}}\biggl(g_{t\phi,r}V_t-g_{tt,r}V_\phi\biggr), \label{eq:energy} \\ 
    &J_z=p_\phi+\frac{S}{2}\sqrt{-\frac{g_{\theta\theta}}{g}}\biggl(g_{t\phi,r}V_\phi-g_{\phi\phi,r}V_t\biggr), \label{eq:angmom}
\end{align}
in the special case of circular equatorial orbits and hold for any stationary, axisymmetric spacetime with reflection symmetry along the equatorial plane (SAR spacetime).

The structure of this article is organized as follows.  Sec.~\ref{sec:ISCOr} revisits the issue of determining the innermost stable circular orbit (ISCO) of a spinning test body in curved spacetime, by suggesting a novel method to calculate it. Such a revision is necessary for contrasting the ISCO frequencies, as functions of the Kerr spin parameter, produced under three different formalisms, the TD, MP and OKS SSCs, analyzed numerically in Sec.~\ref{sec:ISCOcom}. The technical details concerning the power series expansion method, or the analytical algorithm constructed for finding circular equatorial orbits are kept at a minimal extent, since the respective discussions in Paper I were presented in a rather general form, valid for the Kerr spacetime as well. As a result, Sec.~\ref{sec:CEOcom} includes the comparisons of the orbital frequencies for generic circular equatorial orbits, whereas the appropriate shifts between the different SSCs are applied and discussed in Sec.~\ref{sec:CEOcom2} Finally, Sec.~\ref{sec:concl} summarizes the primary findings of this work.   

\textit{Units and notation:} The symbolism of Paper I has been followed throughout the pages of this study, where the central Kerr black hole's mass is denoted by $M$ and the conserved notion of mass on each case ($\mu$ under TD and OKS SSCs, or $m$ for the MP SSC) is identified with the rest mass of the inspiralling body, while $M \gg \mu$ and $M \gg m$ is satisfied respectively. For future reference let us also underline that the Kerr spin parameter is normalized with respect to the central black hole's mass, similar to the circular equatorial orbit radius, i.e.\ $\hat{a}:=\dfrac{a}{M}$ and $\hat{r}:=\dfrac{r}{M}$. As for the rest physical quantities involved in the present work we use the common conventions, introduced in \cite{Timogiannis21, Harms16, Gerakopoulos17}, under which the dimensionless orbital frequency and the test body's spin are given by $\hat{\Omega}:=M \Omega$ as well as $\sigma:=\dfrac{S}{\mu M}$ (TD and OKS SSCs) or $\sigma:=\dfrac{S}{m M}$ (MP SSC), correspondingly. In Sec.~\ref{sec:CEOcom2} though, an alternative notation has been chosen, with $\tilde{\sigma}$ representing the dimensionless spin of the extended test body under the TD SSC. In any case apropos of EMRIs,  $\lvert\sigma\rvert\ll 1$ appears to be a quite reasonable conception. Last but not least, all calculations have been made in geometric units, in which the speed of light and the gravitational constant are set to  $c=G=1$. Moreover, the Riemann tensor is defined as $R^\mu_{\nu\kappa\lambda}=\Gamma^\mu _{\kappa \alpha} \Gamma^\alpha _{\lambda \nu}-\partial_\lambda \Gamma^\mu _{\kappa \nu}-\Gamma^\mu _{\lambda\alpha}\Gamma^\alpha _{\kappa \nu}+\partial_\kappa \Gamma^\mu _{\lambda \nu}$, while the Christoffel symbols are computed from the metric with signature $(-,+,+,+)$, expressed in terms of the standard Boyer-Lindquist coordinates $\{t,r,\theta,\phi\}$.
Einstein's summation convention has been followed, with all indices running from 0 to 3. The Levi-Civita tensor is given by $\epsilon_{\mu\nu\rho\sigma}=\sqrt{-g}\tilde{\epsilon}_{\mu\nu\rho\sigma}$, with the Levi-Civita symbol $\tilde{\epsilon}_{t r \theta \phi}=1$, and $g$ is the determinant of the background metric.

\section{Innermost Stable Circular Orbits} \label{sec:ISCOr}

The innermost stable circular orbit constitutes a stability limit for circular equatorial motion of a particle around a black hole. Namely, orbits with $r>r_{\rm ISCO}$ are stable, while the ones with $r<r_{\rm ISCO}$ are unstable. The techniques implemented to determine the radius of this particular orbit for geodesic motion in a Kerr spacetime span from the important analytical contribution of Bardeen et al.~\cite{Bardeen72} to more recent endeavours like \cite{Stein20}. In the case of a spinning body the first works dealing with the issue can be tracked back to \cite{Rasband73,Tod76}. Since then many papers have dealt with the ISCO issue either numerically \cite{Steinhoff12,Harms16} or analytically \cite{Jefremov15}. In particular, in the latter work a spinning body's ISCO radius has been evaluated in a Taylor expansion form for the general Kerr background, with emphasis given on the Schwarzschild and extreme Kerr examples. 

\begin{figure}[h]
  \graphicspath{{./PhD/}}
 \centering
 \includegraphics[width=0.5\textwidth]{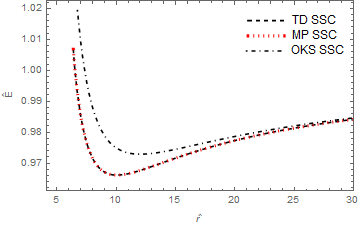} 
   \caption{The plot illustrates the dimensionless energy of an extended spinning test body moving in circular equatorial orbits of different radii, for $\sigma=-0.9$ and $\hat{a}=-0.9$, under the TD, MP and OKS SSCs. The minimum of these functions corresponds to the ISCO radius, which is in agreement with the values and the methods provided in \cite{Harms16_2}. Note also that the curves associated with the TD as well as MP SSCs are nearly indistinguishable, for the whole range of dimensionless radii. }
   \label{fig:ISCOEn}
\end{figure}

\begin{figure}[h]
  \graphicspath{{./PhD/}}
 \centering
    \includegraphics[width=0.5\textwidth]{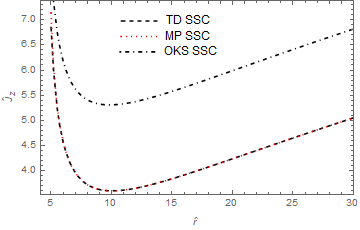}  
 \caption{The panel depicts the z-component of the total angular momentum in appropriate units ($\hat{J}_z=\frac{J_z}{\mu M}$ or $\hat{J}_z=\frac{J_z}{m M}$ based on the SSC chosen to close the MPD equations, see for instance \cite{Timogiannis21, Harms16, Gerakopoulos17}) of a spinning body, as computed under the TD, MP and OKS conditions, versus its distance from a Kerr black hole's center, for $\sigma=-0.9$ and $\hat{a}=-0.9$. Once again, the ISCO radius is detected at the minima of the three curves, whereas the TD and MP prescriptions appear to be very similar.}
 \label{fig:ISCOJz}
\end{figure}

The present section provides an alternative insight to the problem, by using a post-Newtonian analogy, where the ISCO of an equal mass spinning binary is defined in terms of the minimum Bondi binding energy circular orbit (see for instance, \cite{Blanchet13}). Thus, the derivation of the ISCO radius follows from the demand that $\frac{d E}{dr}=0$, or equivalently $\frac{d J_z}{dr}=0$, for the energy and the $z$-component of the total angular momentum of circular equatorial orbits around a Kerr black hole. This can be seen in Figs.~\ref{fig:ISCOEn}, \ref{fig:ISCOJz} for the TD, MP and OKS SSCs. 

Although the proposed procedure lowers the differentiation rank~--when compared to the effective potential method--, its implementation entails a series of major technical setbacks, particularly under the OKS SSC. Consequently, along the following lines we will give: a) a subtle description of the analytical algorithm constructed to determine the ISCO radius, in the case of the Kerr geometry, for each SSC separately; b) numerical results for the TD and MP SSCs.

\subsection{Tulczyjew-Dixon SSC}
 
To find the ISCO radius under the TD SSC one replaces the reference four-vector $V^\nu:=p^\nu/\mu$ in Eqs.~\eqref{eq:energy}, \eqref{eq:angmom}, for the spinning body's energy and z-component of its total angular momentum respectively, a process that yields:
\begin{align}
    E&=\frac{(a M S-\mu r^3)p_t+M S p_\phi}{\mu r^3}, \\
    J_z&=\frac{S (M a^2-r^3)p_t+(\mu r^3+a M S)p_\phi}{\mu r^3}.
\end{align}
In addition, the time and azimuthal covariant components of the four-momentum are substituted by its contravariant counterparts, i.e. $p_\mu=g_{\mu \nu}p^\nu$, that are given by Eqs.~(22) and (23) of Paper I, with the orbital frequency $\Omega_{\pm}$ expressed in terms of the orbital radius, through Eq.~(21) of the above-mentioned paper as well. We wish to underline at this point, that the $+$ sign corresponds to corotation, whereas the $-$ sign is related to counter rotation, with respect to the total angular momentum $J_z$ as in Paper I. Consequently, the desirable result arises, which is that the energy and the total angular momentum along the z-axis have transformed into functions of $r$. The procedure discussed above alludes that the issue of finding the ISCO radius has been addressed for the TD SSC, since it allows the computation of the minima of the functions $E(r;a;S)$ and $J_z(r;a;S)$. The exact expressions of theses functions can be found in \cite{Saijo98,Piovano20} and in the supplemental material \texttt{Mathematica Notebook} \cite{SupMatMP} of the present article. As an example of the novel technique we provide the ISCO radius for different values of the dimensionless spin and Kerr parameter in  Table~\ref{tab:riscoTD}. These values are in agreement with the results found in other works, see, e.g., \cite{Harms16_2,Harms16,Gerakopoulos17}.

\begin{table}[h]
\centering
 \normalsize
 \renewcommand{\arraystretch}{1.5}
    \begin{tabular}{ |c| c| c| c|}
\hline
   \backslashbox{$\sigma$}{$\hat{a}$}&-0.9 & 0.0 & +0.9 \\   \hline 
   -0.9 & 10.0629 & 7.2135 & 3.1472\\ \hline
   -0.5 & 9.5171 & 6.7294 &  2.8411\\ \hline
   -0.1 & 8.8903 & 6.1594 & 2.4294\\ \hline
   +0.1 & 8.5365 & 5.8325 & 2.2155\\ \hline
   +0.5 & 7.7104 & 5.0633 & 1.8726\\ \hline
   +0.9 & 6.6062 & 4.0834 & 1.6632\\ \hline
  \end{tabular}
   \caption[caption]{The location of the last stable orbit $\hat{r}_{\rm ISCO}$ for a spinning body rotating around a Kerr black hole, described by the TD SSC, presented with four-digit accuracy. These values were acquired within the \texttt{Mathematica} environment by determining the minima of the $E(r;a;S)$ function. More technical details are provided in a \texttt{Mathematica Notebook} as supplemental material \cite{SupMatMP} of the present article.}
   \label{tab:riscoTD}
    \end{table}

\subsection{Mathisson-Pirani SSC}

Under the imposition of the MP condition the observer comoves in a reference frame, which coincides with the rest frame of the extended spinning body. Thus, the reference four-vector corresponds to the test body's four-velocity, or in other words $V^\nu:=u^\nu$ and Eqs.~\eqref{eq:energy}, \eqref{eq:angmom} take the form:
\begin{align}
    E&=\frac{M S(u_\phi+a u_t)-r^3 p_t}{r^3}, \\
    J_z&=\frac{S[(M a^2-r^3)u_t+M a u_\phi]+r^3 p_\phi}{r^3}.
\end{align}
Similarly to the TD SSC, $p_t$ as well as $p_\phi$ have to be expressed with the aid of the metric selected to characterize spacetime, $p_\mu=g_{\mu \nu}p^\nu$, while an analogous association, $u_\mu=g_{\mu \nu}u^\nu$, is also taken into consideration. These constraints combined with the definition equation of the orbital frequency, $\Omega=u^\phi/u^t$, and relations (8), (25), (26) of Paper I as well, provide the test body's energy together with the z-component of its total angular momentum as functions of $r$, $\Omega$. For the last step of the derivation the orbital frequency is substituted, by solving Eq.~(27) of Paper I, for the case of the Kerr background. An extensive discussion regarding the physically accepted solutions is presented in \cite{Costa18} for the Schwarzschild black hole limit, the arguments of which can be generalized to the Kerr black hole background as well. Finally, from the proposed analysis one acquires the expressions $E(r;a;S)$ and $J_z(r;a;S)$\footnote{The lengthy expressions of $E(r;a;S)$ and $J_z(r;a;S)$ can be found in the supplemental material \texttt{Mathematica Notebook} \cite{SupMatTD}.}, with the ISCO radius determined by the minima of each function. By implementing this method we were able to derive the ISCO radius for a spinning test body moving in Kerr spacetime, which is in complete accordance with the results presented in \cite{Harms16} and \cite{Gerakopoulos17}.

\begin{table}[h]
\centering
 \normalsize
 \renewcommand{\arraystretch}{1.5}
    \begin{tabular}{ |c| c| c| c|}
\hline
   \backslashbox{$\sigma$}{$\hat{a}$}&-0.9 & 0.0 & +0.9 \\   \hline 
   -0.9 & 10.0617 & 7.2084 & $\times$\\ \hline
   -0.5 & 9.5170 & 6.7290 &  2.8147\\ \hline
   -0.1 & 8.8903 & 6.1594 & 2.4294\\ \hline
   +0.1 & 8.5365 & 5.8325 & 2.2154\\ \hline
   +0.5 & 7.7089 & 5.0575 & 1.8412\\ \hline
   +0.9 & 6.5704 & 3.8774 & $\times$\\ \hline
  \end{tabular}
   \caption[caption]{The Table illustrates the location of the last stable orbit $\hat{r}_{\rm ISCO}$ for a spinning body rotating around a Kerr black hole, described by the MP SSC, given with four-digit accuracy. The $\times$ symbol denotes the failure of the algorithm for large spin values and $\hat{a}>0$. The same pattern is also observed in the effective potential method, see for instance Table I of \cite{Gerakopoulos17}. We employ the \texttt{FindRoot} routine of \texttt{Mathematica} in order to locate the last stable orbits, by limiting the search region to the corresponding values of the TD SSC in Table \ref{tab:riscoTD}. A more thorough analysis is also included in a \texttt{Mathematica Notebook} as supplemental material \cite{SupMatTD}.}
   \label{tab:riscoMP}
    \end{table}

\subsection{Ohashi-Kyrian-Semer\'{a}k SSC}
In the OKS framework we use the standard notation for the reference four-vector $V^\nu$, associated to the observer of the spinning test body. Under this assumption Eqs.~\eqref{eq:energy}, \eqref{eq:angmom} for the energy and the total angular momentum along the z-direction become:
\begin{align}
    E&=\frac{M S(V_\phi+a V_t)-m r^3 u_t}{r^3}, \\
    J_z&=\frac{S[(M a^2-r^3)V_t+M a V_\phi]+m r^3 u_\phi}{r^3}.
\end{align}
In Paper I we presented a novel analytical technique for finding the orbital frequency of a spinning test body, moving in circular equatorial orbits around a central, massive Schwarzschild black hole, under the OKS SSC. In this work we generalize to the case of a Kerr black hole. As a result, the combination of the orbital frequency definition relation $\Omega=u^\phi/u^t$, with Eqs.~(15), (16) of Paper I reads:

\begin{equation} \label{eq:OKS1}
   M\biggl(1-a \Omega\biggr)\biggl(V^t-a V^\phi\biggr)=r^3\Omega V^\phi,
 \end{equation}
   along with the equation:
\begin{widetext}
   \begin{align} \label{eq:OKS2}
   m r^2 u^t\biggl[M\biggl(1-a\Omega\biggr)^2-r^3\Omega^2\biggr]=-3M S a\biggl(1-a \Omega\biggr)\biggl(V^t-a V^\phi\biggr)
   +M S r^2\biggl[2V^\phi\biggl(1-a\Omega\biggr)+\Omega\biggl(V^t-a V^\phi\biggr)\biggr].
\end{align}
\end{widetext}
The emergence of the $\biggl(V^t-a V^\phi\biggr)$ terms in Eqs.~\eqref{eq:OKS1}, \eqref{eq:OKS2} is crucial for the derivation of the polynomial equation satisfied by the extended, spinning test body's orbital frequency. This is achieved by taking advantage of the normalization condition of the reference four-vector $V^\nu$, i.e $V^\nu V_\nu=-1$, apart from relations \eqref{eq:OKS1} and \eqref{eq:OKS2}, which gives: 
\begin{widetext}
\begin{align}  \label{eq:OKSpol}
    &\Omega^6\biggl\{m^2 r^{13}-2M m^2  r^{12}-4 M m^2 a^2  r^{10}+M^2 r^9\biggl(3 m^2 a^2+ S^2 \biggr)+a^2 M^2 r^7\biggl(5m^2 a^2+7 S^2 \biggr)+6a^2 M^3 S^2 r^6+15a^4 M^2 S^2 r^5\nonumber \\
&-2M^3 a^4 r^4\biggl(m^2 a^2-14  S^2\biggr)-a^4 M^2 r^3\biggl[M^2\biggl(m^2 a^2-12 S^2\biggr)-9a^2 S^2\biggr]+30 M^3 a^6  S^2 r^2+28a^6 M^4 S^2 r+8a^6 M^5 S^2\biggr\}\nonumber \\
&+2M a \Omega^5\biggl\{3m^2 r^{10}-3M m^2 r^9-M r^7\biggl(8m^2 a^2+3S^2\biggr)-6M^2 S^2 r^6-12M a^2 S^2 r^5+5a^2 M^2 r^4\biggl(m^2 a^2-8S^2\biggr) \nonumber \\
&+3 M a^2 r^3\biggl[M^2\biggr(m^2 a^2-8S^2\biggr)-3 a^2 S^2 \biggr]-54a^4 M^2 S^2 r^2-68 a^4 M^3 S^2 r-24 a^4 M^4 S^2\biggr\}+M \Omega^4\biggl[-2m^2 r^{10}+3M m^2r^9 \nonumber \\
&+M r^7\biggl(18m^2 a^2-S^2\biggr)+6M^2 S^2 r^6+3M a^2 S^2 r^5-4a^2 M^2 r^4\biggl(5m^2 a^2-18 S^2\biggr)-3a^2 M^3 r^3\biggl(5m^2 a^2-24 S^2\biggr) \nonumber \\
&+132a^4 M^2 S^2 r^2+260a^4 M^3 S^2 r+120 a^4 M^4 S^2\biggr]+2a M^2 \Omega^3\biggl\{-4m^2 r^7+3S^2 r^5+2M r^4\biggl(5m^2 a^2-4S^2\biggr)\nonumber \\
&+r^3\biggl[2M^2\biggl(5m^2 a^2-12S^2\biggr)+9a^2 S^2\biggr] -24M a^2 S^2 r^2-120a^2 M^2 S^2 r-80a^2 M^3 S^2\biggr\}\nonumber \\
&+M^2 \Omega^2\biggl\{m^2 r^7-2M r^4\biggl(5m^2 a^2+2S^2\biggr)-3r^3\biggl[M^2\biggl(5m^2 a^2-4S^2\biggr)+3a^2 S^2\biggr]-18M a^2 S^2 r^2+100a^2 M^2 S^2 r+120a^2 M^3 S^2\biggr\} \nonumber \\
&+2a M^3 \Omega\biggl(m^2 r^4+3M m^2 r^3+6S^2 r^2-4M S^2 r-24M^2 S^2\biggr)-M^4\biggl(m^2 r^3+4 S^2 r-8M S^2\biggr)=0.
\end{align}
\end{widetext}

Just for a brief crosscheck note that Eq.~\eqref{eq:OKSpol} is identical to the corresponding polynomial included in Paper I, when $a=0$. Contrary to the Schwarzschild black hole limit, the presence of the Kerr parameter in the general case renders the sextic equation technically unsolvable. The latter contrast is well understood, if we notice that for the Schwarzschild background spacetime the odd coefficients vanish, and Eq.~\eqref{eq:OKSpol} reduces to a cubic polynomial equation with respect to $\Omega^2$. Thereby, the lack of a $\Omega(r;a;S)$ function suggests that the discussed algorithm of Sec.~2 for finding ISCOs, is not applicable under the OKS SSC. Thus, the effective potential method introduced in \cite{Harms16, Gerakopoulos17} is a necessity rather than a choice for this specific SSC, unless one is able to solve a generic sextic polynomial equation.

\section{ISCO COMPARISONS} \label{sec:ISCOcom}

The notion of ISCO is  important for the study of compact object mechanics, since it divides the equatorial orbits with respect to their stability. Noticeably, ISCO apparently marks the maximum regime of convergence for spinning test bodies' orbital frequencies, among different SSCs, according to our findings in Paper I. The latter fact originates from the dynamically invariant nature of the last stable orbit, which is described in-depth, in our previous work. As a result, we prefer to start the comparisons from the ISCO frequencies, based on the analysis of Sec.~3 of Paper I. With that being said, the discussion of Sec.~\ref{sec:ISCOr} indicates that the numerical manipulation of the problem is unavoidable. Regarding the power series expansion method we recall the orbital frequency form  $\hat{\Omega}=\hat{\Omega}_n \sigma^n+\mathcal{O}\left(\sigma^5\right)$, with $n$ varying from $0$ to $4$. 

\begin{figure}[h]
 \graphicspath{{./PhD/}}
  \centering
    \includegraphics[width=.85\linewidth]{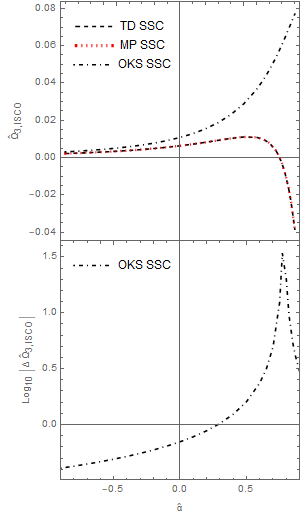}
\caption{The top panel represents the alteration of the $\mathcal{O}(\sigma^3)$-term for the ISCO orbital frequency of a spinning test body, due to the presence of the Kerr parameter (in appropriate units), computed under the three examined SSCs, respectively. Furthermore, the bottom panel illustrates the absolute value of the relative difference of the $\hat{\Omega}_{3,{\rm ISCO}}$ measured in the OKS reference frame, compared to the corresponding term of the TD SSC, on logarithmic scale. The comparison between the MP and the TD SSC has been omitted, since the relative discrepancies appear at higher order.}
\label{fig:sigma3}
\end{figure}
\begin{figure}[h]
   \graphicspath{{./PhD/}}
  \centering
    \includegraphics[width=.85\linewidth]{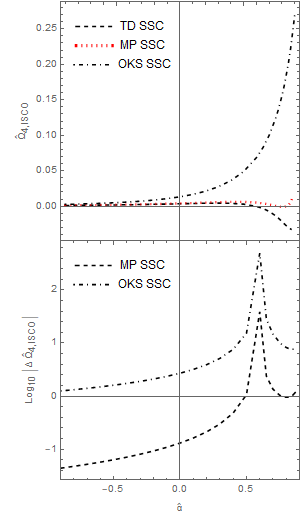}
    \caption{The top panel depicts the $\mathcal{O}(\sigma^4)$-term of the extended test body's orbital frequency at the ISCO radius versus the dimensionless Kerr parameter $\hat{a}$, under the TD, MP and OKS formalisms correspondingly. In addition, the bottom panel demonstrates the absolute value of the relative difference of the $\hat{\Omega}_{4,{\rm ISCO}}$, determined in the MP and OKS reference frames, contrasted to the TD SSC, on logarithmic scale.}
    \label{fig:sigma4}
 \end{figure}   
The expansion coefficients $\hat{\Omega}_n$ for each SSC are computed by substitution in Eqs.~(21), (27) of Paper I as well as Eq.~\eqref{eq:OKSpol} of the present work. Particularly, for the ISCO orbital frequency investigated here, we employ the effective potential method, introduced in \cite{Harms16,Hackmann14}, in order to evaluate numerically the location of the last stable orbit, in terms of a power series expansion. As a consequence, the power series of the ISCO orbital frequency is derived from the power series of the circular equatorial orbit frequency, when the replacement $\hat{r}=\hat{r}_{\rm ISCO}$ takes place.
The results are summarized in Figs.~\ref{fig:sigma3}, \ref{fig:sigma4} where we recover the same trend that governed the Schwarzschild case as well \cite{Timogiannis21}, that is the TD and MP SSCs are identical up to $\mathcal{O}(\sigma^3)$-terms, while the convergence of the OKS SSC is weaker and limited to quadratic spin terms. In an effort to argue quantitatively we established the relative differences $\Delta \hat{\Omega}_{3,\rm ISCO}$ and $\Delta \hat{\Omega}_{4,\rm ISCO}$ given by the relations:   
\begin{align}
 &\Delta \hat{\Omega}_{3,\rm ISCO}=\frac{\hat{\Omega}_{3,\rm OKS, \rm ISCO}-\hat{\Omega}_{3,\rm TD, \rm ISCO}}{\hat{\Omega}_{3,\rm TD, \rm ISCO}},\\
 &\Delta \hat{\Omega}_{4,\rm ISCO}=\frac{\hat{\Omega}_{4,\rm SSC, \rm ISCO}-\hat{\Omega}_{4,\rm TD,\rm ISCO}}{\hat{\Omega}_{4,\rm TD,\rm ISCO}}.
\end{align} 

We use logarithmic scale for the graphs, since the relative discrepancies of the OKS SSC increase for positive values of the Kerr parameter. It is also worth noticing that the cusp appearing in both lower panels of Figs.~\ref{fig:sigma3}, \ref{fig:sigma4} just corresponds to an alteration of sign of the relative difference and therefore, is not related to any singularity. Let us now focus our interest on the frequency of arbitrary circular equatorial orbits (CEOs). 

\section{CEOs COMPARISONS} \label{sec:CEOcom}

In the present Section we implement the power series expansion method, introduced in Paper I, in order to determine the orbital frequency of an extended test body, moving in circular equatorial orbits around a Kerr black hole, under the TD, MP and OKS SSCs respectively. Namely, we wish to test if the fundamental findings of Paper I are affected, with the change of type of the background spacetime. The basic mathematical tools employed for the comparison can be found in Paper I (given in a generalized SAR version), with a quick review also included in Sec.~\ref{sec:ISCOcom}. For the sake of completeness we note that the solution of the system of Eqs.~(21), (27) of Paper I and Eq.~\eqref{eq:OKSpol} of the current work yields a pair $\hat{\Omega}_{\pm}(\hat{r},\hat{a})$, with the upper sign corresponding to prograde orbits, while the lower sign is related to retrograde orbits. The results are listed in Table~\ref{tab:CEOs1}.

\begin{table}[h] 
\centering
\large
  \renewcommand{\arraystretch}{1.5}
  
   \begin{tabular}{ |c| c| c| c |}
\hline

$ \hat{\Omega}_n $ & TD SSC & MP SSC  & OKS SSC\\   \hline

$\mathcal{O}(\sigma^0)$ & $\frac{1}{\hat{a}\pm\sqrt{\hat{r}^3}}$& $\frac{1}{\hat{a}\pm\sqrt{\hat{r}^3}}$
& $\frac{1}{\hat{a}\pm\sqrt{\hat{r}^3}}$ \\ \hline
$\mathcal{O}(\sigma^1)$ &$\frac{3(\pm\hat{a}- \sqrt{\hat{r}})}{2\sqrt{\hat{r}}(\hat{a}\pm \sqrt{\hat{r}^3})^2}$ & $\frac{3(\pm\hat{a}- \sqrt{\hat{r}})}{2\sqrt{\hat{r}}(\hat{a}\pm \sqrt{\hat{r}^3})^2}$  &
$\frac{3(\pm\hat{a}- \sqrt{\hat{r}})}{2\sqrt{\hat{r}}(\hat{a}\pm \sqrt{\hat{r}^3})^2}$ 
\\ \hline
$\mathcal{O}(\sigma^2)$&$\hat{\Omega}_{2,\textrm{TD}}(\hat{r},\hat{a})$& $\hat{\Omega}_{2,\textrm{MP}}(\hat{r},\hat{a})$
& $\hat{\Omega}_{2,\textrm{OKS}}(\hat{r},\hat{a})$ 
\\ \hline
$\mathcal{O}(\sigma^3)$&$\hat{\Omega}_{3,\textrm{TD}}(\hat{r},\hat{a})$
& $\hat{\Omega}_{3,\textrm{MP}}(\hat{r},\hat{a})$
& $\hat{\Omega}_{3,\textrm{OKS}}(\hat{r},\hat{a})$ 
\\ \hline
$\mathcal{O}(\sigma^4)$&$\hat{\Omega}_{4,\textrm{TD}}(\hat{r},\hat{a})$& $\hat{\Omega}_{4,\textrm{MP}}(\hat{r},\hat{a})$
& $\hat{\Omega}_{4,\textrm{OKS}}(\hat{r},\hat{a})$ 
\\ \hline
     \end{tabular}
   \caption[caption]{The power series expansion coefficients for the frequencies $\hat{\Omega}_{\pm}$ of circular equatorial orbits around a central Kerr black hole, for the TD, MP and OKS SSCs.}
\label{tab:CEOs1}
 \end{table}

Table~\ref{tab:CEOs1} provides a first indication concerning the linear agreement of all examined SSCs, without the application of any centroid corrections, since all $\hat{\Omega}_{1\pm}$ terms are identical. The lengthy expressions for the higher order contributions of the expansion are included in Appendix~\ref{sec:app1}, where we can verify that the TD and MP SSCs have a stronger level of convergence, being compatible up to $\mathcal{O}(\sigma^2)$-terms. For an assessment of the produced results note that when $\hat{a}$ vanishes the columns of Table I in Paper I are recovered, while the $\hat{\Omega}_{\pm}$ of Table \ref{tab:CEOs1} are also valid in the geodesic limit. We would like to mention at this point that a similar analysis is given in \cite{Khodagholizadeh20}, where the authors derive quadratic in spin expansions of the orbital frequency of a test body moving in circular equatorial orbits around a Kerr black hole, under the TD and MP SSCs. Although our findings are in accordance with Eq.~(34) of \cite{Khodagholizadeh20}, the authors state that the $\hat{\Omega}_{2\pm}$ for the TD and MP SSCs are different. In fact this is inaccurate, since Eq.~(40) of Ref.~\cite{Khodagholizadeh20} can be further simplified in order to match with our proposed result.  

\section{CENTROIDS' CORRECTIONS} \label{sec:CEOcom2}

The discussion of Secs.~\ref{sec:ISCOr} and \ref{sec:ISCOcom} designates that the maximum convergence of the test body's orbital frequencies under the examined SSCs, is attained at ISCO. For that reason, in the present Section we improve the agreement of the expansions in Table~\ref{tab:CEOs1} by shifting properly from the MP or OKS reference frames to the TD frame. This choice has been made in order to avoid the undesirable consequences correlated with the MP and OKS SSCs, like the notorious helical motion of the MP SSC (for further explanations see Paper I). The mathematical idea behind the centroid's shift can be found in \cite{Kyrian07} and has been employed in Paper I for the Schwarzschild case. In brief, the spin tensor is transformed through the equation:
\begin{equation} \label{eq:shift1}
    \tilde{S}^{\mu\nu}=S^{\mu\nu}+p^\mu \delta z^\nu-p^\nu \delta z^\mu,
\end{equation}
when the centroid's worldline is shifted towards $\tilde{z}^\nu=z^\nu+\delta z^\nu$ with:
\begin{equation} \label{eq:shift2}
    \delta z^\nu=\frac{\tilde{p}_\mu S^{\mu\nu}}{\tilde{\mu}^2},
\end{equation}
and $\tilde{\mu}^2=-\tilde{g}_{\kappa\sigma}p^\kappa p^\sigma$ defined as the dynamical rest mass. We remind at this point that the tilde symbol denotes the quantities measured in the TD reference frame, whereas the rest quantities are computed within the MP or OKS frameworks respectively. Following the main structure of Paper I, we consider the option of radial shifts of the centroid, justified by the analysis which can be found in the Appendix~B of the aforementioned work. Namely, this analysis showed that , other center of mass shifts either lead to unphysical effects or are not practical for calculations.

\subsection{Radial Linear Corrections}

The introduction of the correction in Eq.~\eqref{eq:shift2} can drastically improve the power series convergence, summarized in Table~\ref{tab:CEOs1}. This is achieved by considering a radial shift of the form:
\begin{equation} \label{eq:shift3}
\tilde{r}=r+\delta r,
\end{equation}
with $\delta r$ described by Eq.~\eqref{eq:shift2}. In our first approximation we assume that the spin measure of the test body is preserved under the alteration of SSC, i.e. $\tilde{\sigma}=\sigma$. The expansion of Eq.~\eqref{eq:shift2} in terms of the shift $\delta r$, for the generic Kerr case leads to:
\begin{widetext}
\begin{align} \label{eq:deltar}
\delta r&=\frac{p_t S^{tr}+p_\phi S^{\phi r}}{\mu^2}+\frac{\delta r}{\mu^2(g_{tt}g_{\phi\phi}-{g_{t\phi}}^2)}\biggl\{\biggl[g_{tt,r}\biggl(g_{\phi\phi}p_t-g_{t\phi}p_\phi\biggr)+g_{t\phi,r}\biggl(g_{tt}p_\phi-g_{t\phi}p_t\biggr)\biggr]S^{tr}+\biggl[g_{t\phi,r}\biggl(g_{\phi\phi}p_t-g_{t\phi}p_\phi\biggr) \nonumber\\
&+g_{\phi\phi,r}\biggl(g_{tt}p_\phi-g_{t\phi}p_t\biggr)\biggr]S^{\phi r}+\biggl[\frac{p_t S^{tr}+p_\phi S^{\phi r}}{\mu^2(g_{tt}g_{\phi\phi}-{g_{t\phi}}^2)}\biggr]\biggl[g_{tt,r}\biggl(g_{\phi\phi}p_t-g_{t\phi}p_\phi\biggr)^2+2g_{t\phi,r}\biggl(g_{\phi\phi}p_t-g_{t\phi}p_\phi\biggr)\biggl(g_{tt}p_\phi-g_{t\phi}p_t\biggr) \nonumber\\
&+g_{\phi\phi,r}\biggl(g_{tt}p_\phi-g_{t\phi}p_t\biggr)^2\biggr]\biggr\}+\mathcal{O}(\delta r^2).  
\end{align}
\end{widetext}

The fundamental correction of the position of the centroid arises by neglecting the $\mathcal{O}(\delta r)$-term in the RHS of Eq.~\eqref{eq:deltar}, for the sake of simplicity. In Paper I we delineate the algorithmic process followed to derive $\delta r$ in terms of a power series expansion, in a form that is valid for every SAR spacetime and therefore can be applied for the Kerr metric as well (the interested reader can find a more sophisticated analysis in Section 4.1 of Paper I). As a result, the radial distance in the first column of Table \ref{tab:CEOs1} is adjusted accordingly based on Eq.~\eqref{eq:shift3}, for the TD-MP and TD-OKS pair of SSCs. The results are demonstrated in Tables~\ref{tab:CEOs2},~\ref{tab:CEOs3}. 

\begin{table}[h]
\centering
\large
   \renewcommand{\arraystretch}{1.5}
  
   \begin{tabular}{ |c| c| c|}
\hline

$ \hat{\Omega}_n $ & TD SSC & MP SSC \\   \hline

$\mathcal{O}(\sigma^0)$ & $\frac{1}{\hat{a}\pm\sqrt{\hat{r}^3}}$& $\frac{1}{\hat{a}\pm\sqrt{\hat{r}^3}}$ \\ \hline
$\mathcal{O}(\sigma^1)$ &$\frac{3(\pm\hat{a}- \sqrt{\hat{r}})}{2\sqrt{\hat{r}}(\hat{a}\pm \sqrt{\hat{r}^3})^2}$ & $\frac{3(\pm\hat{a}- \sqrt{\hat{r}})}{2\sqrt{\hat{r}}(\hat{a}\pm \sqrt{\hat{r}^3})^2}$  \\ \hline
$\mathcal{O}(\sigma^2)$&$\hat{\Omega}_{2,\textrm{MP}}(\hat{r},\hat{a})$& $\hat{\Omega}_{2,\textrm{MP}}(\hat{r},\hat{a})$\\ \hline
$\mathcal{O}(\sigma^3)$&$\hat{\Omega}_{3,\textrm{MP}}(\hat{r},\hat{a})$
& $\hat{\Omega}_{3,\textrm{MP}}(\hat{r},\hat{a})$\\ \hline
$\mathcal{O}(\sigma^4)$&$\hat{\Omega}_{4,\textrm{TD}}^{'}(\hat{r},\hat{a})$& $\hat{\Omega}_{4,\textrm{MP}}(\hat{r},\hat{a})$ \\ \hline
     \end{tabular}
   \caption[caption]{The power series expansion coefficients for the frequencies $\hat{\Omega}_{\pm}$ of circular equatorial orbits around a central Kerr black hole, for the TD and MP pair of SSCs, when $\tilde{r}\neq r$.}
   \label{tab:CEOs2}
\end{table}

\begin{table}[h] 
\centering
\large
   \renewcommand{\arraystretch}{1.5}
  
   \begin{tabular}{ |c| c| c|}
\hline

$ \hat{\Omega}_n $ & TD SSC & OKS SSC \\   \hline

$\mathcal{O}(\sigma^0)$ & $\frac{1}{\hat{a}\pm\sqrt{\hat{r}^3}}$& $\frac{1}{\hat{a}\pm\sqrt{\hat{r}^3}}$ \\ \hline
$\mathcal{O}(\sigma^1)$ &$\frac{3(\pm\hat{a}- \sqrt{\hat{r}})}{2\sqrt{\hat{r}}(\hat{a}\pm \sqrt{\hat{r}^3})^2}$ & $\frac{3(\pm\hat{a}- \sqrt{\hat{r}})}{2\sqrt{\hat{r}}(\hat{a}\pm \sqrt{\hat{r}^3})^2}$  \\ \hline
$\mathcal{O}(\sigma^2)$&$\hat{\Omega}_{2,\textrm{OKS}}(\hat{r},\hat{a})$& $\hat{\Omega}_{2,\textrm{OKS}}(\hat{r},\hat{a})$\\ \hline
$\mathcal{O}(\sigma^3)$&$\hat{\Omega}_{3,\textrm{TD}}^{'}(\hat{r},\hat{a})$
& $\hat{\Omega}_{3,\textrm{OKS}}(\hat{r},\hat{a})$\\ \hline
$\mathcal{O}(\sigma^4)$&$\hat{\Omega}_{4,\textrm{TD}}^{''}(\hat{r},\hat{a})$& $\hat{\Omega}_{4,\textrm{OKS}}(\hat{r},\hat{a})$ \\ \hline
     \end{tabular}
   \caption[caption]{The power series expansion coefficients for the frequencies $\hat{\Omega}_{\pm}$ of circular equatorial orbits around a central Kerr black hole, for the TD and OKS pair of SSCs, when $\tilde{r}\neq r$.}
   \label{tab:CEOs3}
   \end{table}

Tables~\ref{tab:CEOs2},~\ref{tab:CEOs3} confirm the pattern showed for the Schwarzschild background spacetime, which was discussed in Paper I. Namely, the imposition of the centroid's shift fixes the discrepancies in the $\mathcal{O}(\sigma^3)$-terms of the TD-MP pair of SSCs and removes the dissimilarity in the $\mathcal{O}(\sigma^2)$-terms for the TD-OKS pair of conditions. The inclusion of the complete expression
of $\delta r$ from Eq.~\eqref{eq:deltar} leads to the quantities $\hat{\Omega}_{4,\textrm{TD}}^{'''}(\hat{r},\hat{a})$ (TD-MP pair) as well as $\hat{\Omega}_{3,\textrm{TD}}^{''}(\hat{r},\hat{a})$ (TD-OKS pair)\footnote{The functions $\hat{\Omega}_{4,\textrm{TD}}^{'''}(\hat{r},\hat{a})$ and $\hat{\Omega}_{3,\textrm{TD}}^{''}(\hat{r},\hat{a})$ are presented in the Appendix \ref{sec:app1}}, but does not further improve the degree of convergence among the examined SSCs. 

\subsection{Spin Measure Corrections}

An exhaustive analysis on the behaviour of the centroid of a spinning test body should take into account the firmly SSC-dependent nature of the spin measure. In other words, the transition between different SSCs alters the representative worldline with respect to which the moments are evaluated, a process that affects the measure of the spin itself. In the present section we consider the case $\tilde{\sigma}\neq \sigma$ and we produce the corresponding power series expansions for the $\hat{\Omega}_\pm$ orbital frequencies. The latter is achieved by combining Eqs.~\eqref{eq:spin_m} and \eqref{eq:shift1} for a spinning test body moving in circular equatorial orbits around a supermassive Kerr black hole. Recall that for that kind of orbits the only non-vanishing components of the spin tensor are $S^{tr}=-S^{rt}$ along with $S^{r\phi}=-S^{\phi r}$. The described expansion procedure yields: %
\begin{align} \label{eq:spinmeasure}
  \tilde{S}^2 &=S^2+\delta r \biggl\{g_{rr}\biggl[g_{\phi\phi,r}\biggl(S^{r\phi}\biggr)^2-2g_{t\phi,r}S^{r\phi}S^{tr} \nonumber \\
 &+g_{tt,r}\biggl(S^{tr}\biggr)^2 +2\biggl(p_t S^{tr}-p_\phi S^{r\phi}\biggr)\biggr]+\frac{S^2 g_{rr,r}}{g_{rr}}\biggr\}\nonumber\\
 &+\mathcal{O}(\delta r^2). 
\end{align}
For the next step of the derivation we shall make all quantities appear in Eq.~\eqref{eq:spinmeasure} dimensionless, which is a subject that is discussed in the following paragraphs, for the pair of TD-MP and TD-OKS SSCs respectively.

\subsubsection{TD-MP Relation}

In order to acquire a $\tilde{\sigma}=f(\sigma)$ relation for the shift from the MP to the TD centroid we should divide both sides of Eq.~\eqref{eq:spinmeasure} by $\tilde{\mu}^2 M^2$. We also notice that the linear approximation in $\delta r$ of the spinning body's inverse square of the dynamical rest mass reads: 
\begin{align} \label{eq:AMAN}   
    &\frac{1}{\tilde{\mu}^2}=\frac{1}{\mu^2}\biggl\{1+\frac{\delta r}{\mu^2(g_{tt}g_{\phi\phi}-{g_{t\phi}}^2)^2}\biggl[g_{tt,r}\biggl(g_{\phi\phi}p_t-g_{t\phi}p_\phi\biggr)^2\nonumber \\ 
    &+2g_{t\phi,r}\biggl(g_{\phi\phi}p_t-g_{t\phi}p_\phi\biggr)\biggl(g_{tt}p_\phi-g_{t\phi}p_t\biggr)\nonumber\\
    &+g_{\phi\phi,r}\biggl(g_{tt}p_\phi-g_{t\phi}p_t\biggr)^2\biggr]\biggr\}+\mathcal{O}(\delta r^2).
\end{align}
Since $\tilde{\sigma}$ and $\sigma$ are not identically defined, i.e. $\sigma=\frac{S}{m M}$, while $\tilde{\sigma}=\frac{\tilde{S}}{\tilde{\mu}M}$ one needs to correlate the dynamical rest mass $\mu$ with the kinematical rest mass $m$, both measured in the MP reference frame. Such a link is provided in \cite{Costa18} and more precisely:
\begin{equation*}
 \mu^2=m^2+\frac{S^{\alpha\kappa}S_{\kappa\beta}p^\beta p_\alpha}{S^2}=m^2-\frac{g_{rr}(p_t S^{tr}-p_\phi S^{r \phi})^2}{S^2},
\end{equation*}
which coincides with the expression derived in Eq.~(44) of \cite{Timogiannis21} for the Schwarzschild background spacetime. Consequently, the desirable relation between $\tilde{\sigma}$ and $\sigma$ in terms of a power series expansion takes the form: 
\begin{widetext}
\begin{align} \label{eq:firstsigma}
    \tilde{\sigma}-\sigma&=\frac{3(\hat{a}\mp\sqrt{\hat{r}})[4\hat{a}^4\mp16\hat{a}^3\sqrt{\hat{r}}+16\hat{a}^2\hat{r}\pm4\hat{a}^3\sqrt{\hat{r}^3}-4\hat{a}^2\hat{r}^2\mp2\hat{a}(\hat{a}^2+4)\sqrt{\hat{r}^5}]\sigma^4}{\hat{r}^8[2\hat{a}\pm  \sqrt{\hat{r}}(\hat{r}-3)]^2} \nonumber \\
    &+\frac{3(\hat{a}\mp\sqrt{\hat{r}})[3\hat{a}^2\hat{r}^3\pm8\hat{a}\sqrt{\hat{r}^7}-(\hat{a}^2+6)\hat{r}^4\mp2\hat{a}\sqrt{\hat{r}^9}+5\hat{r}^5-\hat{r}^6]\sigma^4}{\hat{r}^8[2\hat{a}\pm  \sqrt{\hat{r}}(\hat{r}-3)]^2}+\mathcal{O}(\sigma^5),
\end{align}
\end{widetext}
which reduces to Eq.~(47) of Paper I in the Schwarzschild black hole limit. Eq.~\eqref{eq:firstsigma} implies that the orbital frequency expansion coefficients that satisfy the inequality $\hat{\Omega}_{n\pm}\leqslant\mathcal{O}(\sigma^3)$ are not influenced from the alteration of the spin measure. By substituting the function $\tilde{\sigma}=f(\sigma)$ to the $\hat{\Omega}_{4,\textrm{TD}}^{'''}(\hat{r},\hat{a})$ we extract the quantity $\hat{\Omega}_{4,\textrm{TD}}^{''''}(\hat{r},\hat{a})$ (included in the Appendix \ref{sec:app1}), which remains indifferent to the $\hat{\Omega}_{4,\textrm{MP}}(\hat{r},\hat{a})$ term. The latter fact provides a concrete indication that the convergence of the frequency power series cannot be further improved.   

\subsubsection{TD-OKS Relation}
The constancy of the dynamical rest mass under the OKS SSC simplifies drastically the process of deriving a $\tilde{\sigma}=h(\sigma)$ correlation, which governs the transition from the OKS frame of reference to the TD frame. The division of both hands of Eq.~\eqref{eq:spinmeasure} by $\tilde{\mu}^2M^2$, combined with Eq.~\eqref{eq:AMAN} implies that:
\begin{widetext}
\begin{align} \label{eq:spinoks}
   \tilde{\sigma}^2-\sigma^2&=\delta r\Biggl\{g_{rr}\biggl[g_{\phi\phi,r}\biggl(\sigma^{r\phi}\biggr)^2+2g_{t\phi,r}\sigma^{r\phi}\sigma^{tr}+g_{tt,r}\biggl(\sigma^{tr}\biggr)^2+\frac{2}{\mu M}\biggl(p_t \sigma^{tr}-p_\phi \sigma^{r\phi}\biggr)\biggr] \nonumber \\
   &+\sigma^2\biggl\{\frac{g_{rr,r}}{g_{rr}}
    +\frac{1}{\mu^2(g_{tt}g_{\phi\phi}-{g_{t\phi}}^2)^2}\biggl[g_{tt,r}\biggl(g_{\phi\phi}p_t-g_{t\phi}p_\phi\biggr)^2+2g_{t\phi,r}\biggl(g_{\phi\phi}p_t-g_{t\phi}p_\phi\biggr)\nonumber \\
    &\times\biggl(g_{tt}p_\phi-g_{t\phi}p_t\biggr)+g_{\phi\phi,r}\biggl(g_{tt}p_\phi-g_{t\phi}p_t\biggr)^2\biggr]\biggr\}\Biggr\} 
   +\mathcal{O}(\delta r^2),
\end{align}
\end{widetext}
where we introduced the normalized (but not necessarily dimensionless) spin tensor $\sigma^{\kappa\nu}=\dfrac{S^{\kappa\nu}}{\mu M}$, for the sake of brevity. The complicated expression in Eq.~\eqref{eq:spinoks} takes a more compact form, when one is limited in the case of circular equatorial orbits and more specifically:
\begin{widetext}
\begin{align} \label{eq:secondsigma}
    \tilde{\sigma}-\sigma&=\frac{3(\hat{a}\mp\sqrt{\hat{r}})[\pm4\hat{a}^4-16\hat{a}^3\sqrt{\hat{r}}\pm16\hat{a}^2\hat{r}+4\hat{a}^3\sqrt{\hat{r}^3}\mp4\hat{a}^2\hat{r}^2+\hat{a}(\hat{a}^2-8)\sqrt{\hat{r}^5}]\sigma^3}{\sqrt{\hat{r}^{13}}[2\hat{a}\pm  \sqrt{\hat{r}}(\hat{r}-3)]^2} \nonumber \\
    &+\frac{3(\hat{a}\mp\sqrt{\hat{r}})(2\hat{a}\sqrt{\hat{r}^7}\mp\hat{a}^2\hat{r}^4+\hat{a}\sqrt{\hat{r}^9}\pm2\hat{r}^5\mp\hat{r}^6)\sigma^3}{\sqrt{\hat{r}^{13}}[2\hat{a}\pm  \sqrt{\hat{r}}(\hat{r}-3)]^2}+\mathcal{O}(\sigma^4).
\end{align}
\end{widetext}
In correspondence with the former set of SSCs we exploit the $\tilde{\sigma}=h(\sigma)$ relation in order to produce the $\hat{\Omega}_{3,\textrm{TD}}^{'''}(\hat{r},\hat{a})$ function from the $\hat{\Omega}_{3,\textrm{TD}}^{''}(\hat{r},\hat{a})$ function (look in the Appendix \ref{sec:app1} for the full expressions). It is also clear from the form of Eq.~\eqref{eq:secondsigma} that it applies modifications only to the cubic or higher contributions, but the gap between the TD and OKS SSCs remains unbridgeable.

\section{CONCLUSIONS} \label{sec:concl}
In the present article we continue the investigation of the equivalence of the Mathisson-Papapetrou-Dixon equations under different spin supplementary conditions, a project that started in \cite{Timogiannis21}. More specifically, we examine the orbital frequencies produced by an extended spinning body moving in circular orbits around the equatorial plane of a supermassive Kerr black hole, under the Tulczyjew-Dixon, the Mathisson-Pirani and the Ohashi-Kyrian-Semer\'{a}k spin conditions. For this reason, we exploit the general framework, which has been founded in \cite{Timogiannis21} for an arbitrary, stationary, axisymmetric spacetime with reflection symmetry. The results of the frequencies comparison summarized in Table \ref{tab:CEOs1} indicate that the aforementioned spin supplementary conditions converge up to linear order in spin, while the Tulczyjew-Dixon and the Mathisson-Pirani SSCs appear to have a stronger level of convergence. It should be stressed that throughout our work only the non-helical Mathisson-Pirani centroid was studied and the obtained results hold only for this centroid choice. The introduction of the centroid position corrections can adequately improve the agreement by one order of spin, as it is clear from Tables \ref{tab:CEOs2} and \ref{tab:CEOs3}. Hence, the central conclusion of \cite{Timogiannis21} for the Schwarzschild spacetime is also valid for the more general Kerr background, that is the examined spin supplementary conditions are in accordance up to quadratic spin terms.

The ISCO frequencies play a significant role in understanding the physical reason behind the observed discrepancies among the Tulczyjew-Dixon, the Mathisson-Pirani and the Ohashi-Kyrian-Semer\'{a}k spin supplementary conditions. As a stability limit the innermost stable circular orbit marks the maximum level of convergence among the orbital frequency expansions within the pole-dipole regime. The latter claim becomes more apparent in Figs.~\ref{fig:sigma3} and \ref{fig:sigma4} where we confirm  that the Tulczyjew-Dixon compared to the Mathisson-Pirani spin supplementary condition agree up to cubic spin terms, whereas the Tulczyjew-Dixon and the Ohashi-Kyrian-Semer\'{a}k conditions diverge more rapidly. In the process of performing the frequency comparison between the SSCs at ISCO, we were able to formulate a novel method for finding the radial position of these marginally stable circular orbits. Even if there should be, in principle,  a way to provide analytical expressions in the case of the TD and MP SSCs, we restrict our confirmation on numerical results. The reason behind this decision are extremely long and complicated analytical formulas. Actually, in the case of OKS SSC the problem of determining the ISCO using this novel method is more fundamental, since one has to obtain analytically the roots of a sextic polynomial equation.

\begin{acknowledgments}
G.L.G. has been supported by the fellowship Lumina Quaeruntur No. LQ100032102 of the Czech Academy of Sciences.
\end{acknowledgments}

\appendix
\section{Expansion Coefficients} \label{sec:app1}
The present section of the article contains large expressions related to the orbital frequency power series expansions, demonstrated in Secs.~\ref{sec:CEOcom} and \ref{sec:CEOcom2}.
\begin{widetext}
\begin{align*}
\hat{\Omega}_{2,\textrm{TD}}(\hat{r},\hat{a}) &=\frac{3(\hat{a}\mp\sqrt{\hat{r}})(\mp9\hat{a}^2-\hat{a}\sqrt{\hat{r}}-3\hat{a}\sqrt{\hat{r}^3}\mp7\hat{r}^2)}{8\sqrt{\hat{r}^5}(\hat{a}\pm\sqrt{\hat{r}^3})^3}, \\
\hat{\Omega}_{2,\textrm{MP}}(\hat{r},\hat{a})&=\frac{3(\hat{a}\mp\sqrt{\hat{r}})(\mp9\hat{a}^2-\hat{a}\sqrt{\hat{r}}-3\hat{a}\sqrt{\hat{r}^3}\mp7\hat{r}^2)}{8\sqrt{\hat{r}^5}(\hat{a}\pm\sqrt{\hat{r}^3})^3},\\
\hat{\Omega}_{2,\textrm{OKS}}(\hat{r},\hat{a})&=\frac{3(\hat{a}\mp\sqrt{\hat{r}})(\mp6\hat{a}^3+25\hat{a}^2\sqrt{\hat{r}}\mp21\hat{a}\hat{r}-3\hat{a}^2\sqrt{\hat{r}^3}\pm6\hat{a}\hat{r}^2-3\sqrt{\hat{r}^5}\mp3\hat{a}\hat{r}^3+5\sqrt{\hat{r}^7})}{8\sqrt{\hat{r}^5}(\hat{a}\pm\sqrt{\hat{r}^3})^3[2\hat{a}\pm\sqrt{\hat{r}}(\hat{r}-3)]}, \\
\hat{\Omega}_{3,\textrm{TD}}(\hat{r},\hat{a})&=\frac{3(\hat{a}\mp\sqrt{\hat{r}})(\pm45\hat{a}^4-3\hat{a}^3\sqrt{\hat{r}}\mp16\hat{a}^2\hat{r}+36\hat{a}^3\sqrt{\hat{r}^3}\pm42\hat{a}^2\hat{r}^2-26\hat{a}\sqrt{\hat{r}^5}\pm9\hat{a}^2\hat{r}^3+9\hat{a}\sqrt{\hat{r}^7}\pm8\hat{r}^4)}{16\sqrt{\hat{r}^9}(\hat{a}\pm\sqrt{\hat{r}^3})^4},    \\
 \hat{\Omega}_{3,\textrm{MP}}(\hat{r},\hat{a})&=\frac{3(\hat{a}\mp\sqrt{\hat{r}})[\pm90\hat{a}^5-117\hat{a}^4\sqrt{\hat{r}}\mp23\hat{a}^3\hat{r}+117\hat{a}^4\sqrt{\hat{r}^3}\pm21\hat{a}^3\hat{r}^2-170\hat{a}^2\sqrt{\hat{r}^5}\pm18\hat{a}\hat{r}^3(3\hat{a}^2-1)] }{16\sqrt{\hat{r}^9}(\hat{a}\pm\sqrt{\hat{r}^3})^4[2\hat{a}\pm\sqrt{\hat{r}}(\hat{r}-3)]}\\
 &+\frac{3(\hat{a}\mp\sqrt{\hat{r}})[57\hat{a}^2\sqrt{\hat{r}^7}\pm11\hat{a}\hat{r}^4+9(\hat{a}^2-8)\sqrt{\hat{r}^9}\pm9\hat{a}\hat{r}^5+32\sqrt{\hat{r}^{11}}]}{16\sqrt{\hat{r}^9}(\hat{a}\pm\sqrt{\hat{r}^3})^4[2\hat{a}\pm\sqrt{\hat{r}}(\hat{r}-3)]}, \\
 \hat{\Omega}_{3,\textrm{OKS}}(\hat{r},\hat{a})&=\frac{3(\hat{a}\mp\sqrt{\hat{r}})[-96\hat{a}^5\pm437\hat{a}^4\sqrt{\hat{r}}-3\hat{a}^3\hat{r}(209+12\hat{a}^2)\mp6\hat{a}^2(-48+\hat{a}^2)\sqrt{\hat{r}^3}+318\hat{a}^3\hat{r}^2\mp3\hat{a}^2(158+9\hat{a}^2)\sqrt{\hat{r}^5}]}{16\hat{r}^4(\hat{a}\pm\sqrt{\hat{r}^3})^4[2\hat{a}\pm\sqrt{\hat{r}}(\hat{r}-3)]^2} \\
 &+\frac{3(\hat{a}\mp\sqrt{\hat{r}})[-99\hat{a}\hat{r}^3(-2+\hat{a}^2)\pm413\hat{a}^2\sqrt{\hat{r}^7}-363\hat{a}\hat{r}^4\pm12(6-7\hat{a}^2)\sqrt{\hat{r}^9}+168\hat{a}\hat{r}^5]}{16\hat{r}^4(\hat{a}\pm\sqrt{\hat{r}^3})^4[2\hat{a}\pm\sqrt{\hat{r}}(\hat{r}-3)]^2}\\
 &+\frac{3(\hat{a}\mp\sqrt{\hat{r}})[\pm3(-28+3\hat{a}^2)\sqrt{\hat{r}^{11}}-39\hat{a}\hat{r}^6\pm32\sqrt{\hat{r}^{13}}]}{16\hat{r}^4(\hat{a}\pm\sqrt{\hat{r}^3})^4[2\hat{a}\pm\sqrt{\hat{r}}(\hat{r}-3)]^2},\\
 \hat{\Omega}_{4,\textrm{TD}}(\hat{r},\hat{a})&=\frac{3(\hat{a}\mp\sqrt{\hat{r}})[\mp945\hat{a}^6+207\hat{a}^5\sqrt{\hat{r}}\pm573\hat{a}^4\hat{r}-\hat{a}^3(67+1269\hat{a}^2)\sqrt{\hat{r}^3}\mp909\hat{a}^4\hat{r}^2+1305\hat{a}^3\sqrt{\hat{r}^5}]}{128\sqrt{\hat{r}^{13}}(\hat{a}\pm\sqrt{\hat{r}^3})^5}\\
 &+\frac{3(\hat{a}\mp\sqrt{\hat{r}})[\mp3\hat{a}^2\hat{r}^3(225\hat{a}^2-59)-603\hat{a}^3\sqrt{\hat{r}^7}\pm135\hat{a}^2\hat{r}^4-3\hat{a}(-149+45\hat{a}^2)\sqrt{\hat{r}^9}\mp135\hat{a}^2\hat{r}^5+51\hat{a}\sqrt{\hat{r}^{11}}\mp13\hat{r}^6]}{128\sqrt{\hat{r}^{13}}(\hat{a}\pm\sqrt{\hat{r}^3})^5},\\
 \hat{\Omega}_{4,\textrm{MP}}(\hat{r},\hat{a})&=\frac{3(\hat{a}\mp\sqrt{\hat{r}})[\mp3780\hat{a}^8+9864\hat{a}^7\sqrt{\hat{r}}\mp4761\hat{a}^6\hat{r}-\hat{a}^5(1921+8856\hat{a}^2)\sqrt{\hat{r}^3})]}{128\sqrt{\hat{r}^{13}}(\hat{a}\pm\sqrt{\hat{r}^3})^5[2\hat{a}\pm\sqrt{\hat{r}}(\hat{r}-3)]^2}\\
 &+\frac{3(\hat{a}\mp\sqrt{\hat{r}})[\pm3\hat{a}^4\hat{r}^2(-317+3822\hat{a}^2)+9\hat{a}^3(61+1429\hat{a}^2)\sqrt{\hat{r}^5}\mp\hat{a}^4\hat{r}^3(12823+8721\hat{a}^2)]}{128\sqrt{\hat{r}^{13}}(\hat{a}\pm\sqrt{\hat{r}^3})^5[2\hat{a}\pm\sqrt{\hat{r}}(\hat{r}-3)]^2}\\
 &+\frac{3(\hat{a}\mp\sqrt{\hat{r}})[3\hat{a}^3(1083\hat{a}^2-1747)\sqrt{\hat{r}^7}\pm9\hat{a}^2\hat{r}^4(177+1492\hat{a}^2)-\hat{a}^3(4509\hat{a}^2-5888)\sqrt{\hat{r}^9}\mp9\hat{a}^2\hat{r}^5(1379+87\hat{a}^2)]}{128\sqrt{\hat{r}^{13}}(\hat{a}\pm\sqrt{\hat{r}^3})^5[2\hat{a}\pm\sqrt{\hat{r}}(\hat{r}-3)]^2}\\
 &+\frac{3(\hat{a}\mp\sqrt{\hat{r}})[9\hat{a}(63+152\hat{a}^2)\sqrt{\hat{r}^{11}}\mp\hat{a}^2\hat{r}^6(1215\hat{a}^2-6668)-3\hat{a}(207\hat{a}^2-559)\sqrt{\hat{r}^{13}}\mp9\hat{r}^7(269+107\hat{a}^2)]}{128\sqrt{\hat{r}^{13}}(\hat{a}\pm\sqrt{\hat{r}^3})^5[2\hat{a}\pm\sqrt{\hat{r}}(\hat{r}-3)]^2}\\
     &+\frac{3(\hat{a}\mp\sqrt{\hat{r}})[-\hat{a}(871+135\hat{a}^2)\sqrt{\hat{r}^{15}}\mp15\hat{r}^8(-178+9\hat{a}^2)-45\hat{a}\sqrt{\hat{r}^{17}}\mp685\hat{r}^9]}{128\sqrt{\hat{r}^{13}}(\hat{a}\pm\sqrt{\hat{r}^3})^5[2\hat{a}\pm\sqrt{\hat{r}}(\hat{r}-3)]^2},\\
 \hat{\Omega}_{4,\textrm{OKS}}(\hat{r},\hat{a})&=\frac{3(\hat{a}\mp\sqrt{\hat{r}})[\pm216\hat{a}^9-756\hat{a}^8\sqrt{\hat{r}}\mp4842\hat{a}^7\hat{r}+\hat{a}^6(30385+1404\hat{a}^2)\sqrt{\hat{r}^3}\mp9\hat{a}^5\hat{r}^2(6747+940\hat{a}^2)]}{128\sqrt{\hat{r}^{13}}(\hat{a}\pm\sqrt{\hat{r}^3})^5[2\hat{a}\pm\sqrt{\hat{r}}(\hat{r}-3)]^3}\\
 &+\frac{3(\hat{a}\mp\sqrt{\hat{r}})[9\hat{a}^4(5831+939\hat{a}^2)\sqrt{\hat{r}^5}\pm9\hat{a}^3\hat{r}^3(-1863+3538\hat{a}^2+294\hat{a}^4)-9\hat{a}^4(9316+1281\hat{a}^2)\sqrt{\hat{r}^7}]}{128\sqrt{\hat{r}^{13}}(\hat{a}\pm\sqrt{\hat{r}^3})^5[2\hat{a}\pm\sqrt{\hat{r}}(\hat{r}-3)]^3}\\
 &+\frac{3(\hat{a}\mp\sqrt{\hat{r}})[\mp18\hat{a}^3\hat{r}^4(-4013+51\hat{a}^2)+3\hat{a}^2(-7209+21575\hat{a}^2+999\hat{a}^4)\sqrt{\hat{r}^9}\mp9\hat{a}^3\hat{r}^5(12007+894\hat{a}^2)]}{128\sqrt{\hat{r}^{13}}(\hat{a}\pm\sqrt{\hat{r}^3})^5[2\hat{a}\pm\sqrt{\hat{r}}(\hat{r}-3)]^3}\\
 &+\frac{3(\hat{a}\mp\sqrt{\hat{r}})[-9\hat{a}^2(-7392+1399\hat{a}^2)\sqrt{\hat{r}^{11}}\pm9\hat{a}\hat{r}^6(-1485+6784\hat{a}^2+189\hat{a}^4)-9\hat{a}^2(7757+68\hat{a}^2)\sqrt{\hat{r}^{13}}]}{128\sqrt{\hat{r}^{13}}(\hat{a}\pm\sqrt{\hat{r}^3})^5[2\hat{a}\pm\sqrt{\hat{r}}(\hat{r}-3)]^3}\\
 &+\frac{3(\hat{a}\mp\sqrt{\hat{r}})[\mp9\hat{a}\hat{r}^7(-3230+1951\hat{a}^2)+3(-891+12181\hat{a}^2+81\hat{a}^4)\sqrt{\hat{r}^{15}}\pm18\hat{a}\hat{r}^8(-1429+113\hat{a}^2)]}{128\sqrt{\hat{r}^{13}}(\hat{a}\pm\sqrt{\hat{r}^3})^5[2\hat{a}\pm\sqrt{\hat{r}}(\hat{r}-3)]^3}\\
 &+\frac{3(\hat{a}\mp\sqrt{\hat{r}})[-9(-625+
 968\hat{a}^2)\sqrt{\hat{r}^{17}}\mp45\hat{a}\hat{r}^9(-230+3\hat{a}^2)+9(-435+113\hat{a}^2)\sqrt{\hat{r}^{19}}\mp1845\hat{a}\hat{r}^{10}+971\sqrt{\hat{r}^{21}}]}{128\sqrt{\hat{r}^{13}}(\hat{a}\pm\sqrt{\hat{r}^3})^5[2\hat{a}\pm\sqrt{\hat{r}}(\hat{r}-3)]^3}.
 \end{align*}
\end{widetext} \newpage

\begin{widetext}
\begin{align*}
    \hat{\Omega}_{4,\textrm{TD}}^{'}(\hat{r},\hat{a})&=\frac{3(\hat{a}\mp\sqrt{\hat{r}})[\mp3780\hat{a}^8+11784\hat{a}^7\sqrt{\hat{r}}\mp8793\hat{a}^6\hat{r}-41\hat{a}^5(89+216\hat{a}^2)\sqrt{\hat{r}^3})]}{128\sqrt{\hat{r}^{13}}(\hat{a}\pm\sqrt{\hat{r}^3})^5[2\hat{a}\pm\sqrt{\hat{r}}(\hat{r}-3)]^2}\\
 &+\frac{3(\hat{a}\mp\sqrt{\hat{r}})[\pm21\hat{a}^4\hat{r}^2(293+866\hat{a}^2)+3\hat{a}^3(-777+575\hat{a}^2)\sqrt{\hat{r}^5}\mp\hat{a}^4\hat{r}^3(22807+8721\hat{a}^2)]}{128\sqrt{\hat{r}^{13}}(\hat{a}\pm\sqrt{\hat{r}^3})^5[2\hat{a}\pm\sqrt{\hat{r}}(\hat{r}-3)]^2}\\
 &+\frac{3(\hat{a}\mp\sqrt{\hat{r}})[3\hat{a}^3(3963\hat{a}^2+6061)\sqrt{\hat{r}^7}\pm3\hat{a}^2\hat{r}^4(1660\hat{a}^2-2349)-\hat{a}^3(4509\hat{a}^2+14080)\sqrt{\hat{r}^9}\pm3\hat{a}^2\hat{r}^5(5079+1339\hat{a}^2)]}{128\sqrt{\hat{r}^{13}}(\hat{a}\pm\sqrt{\hat{r}^3})^5[2\hat{a}\pm\sqrt{\hat{r}}(\hat{r}-3)]^2}\\
 &+\frac{3(\hat{a}\mp\sqrt{\hat{r}})[9\hat{a}(-897+280\hat{a}^2)\sqrt{\hat{r}^{11}}\mp\hat{a}^2\hat{r}^6(1215\hat{a}^2+10612)+3\hat{a}(113\hat{a}^2+5039)\sqrt{\hat{r}^{13}}\pm3\hat{r}^7(-1767+767\hat{a}^2)]}{128\sqrt{\hat{r}^{13}}(\hat{a}\pm\sqrt{\hat{r}^3})^5[2\hat{a}\pm\sqrt{\hat{r}}(\hat{r}-3)]^2}\\
     &+\frac{3(\hat{a}\mp\sqrt{\hat{r}})[-\hat{a}(6823+135\hat{a}^2)\sqrt{\hat{r}^{15}}\mp3\hat{r}^8(-1594+45\hat{a}^2)+723\hat{a}\sqrt{\hat{r}^{17}}\mp1069\hat{r}^9]}{128\sqrt{\hat{r}^{13}}(\hat{a}\pm\sqrt{\hat{r}^3})^5[2\hat{a}\pm\sqrt{\hat{r}}(\hat{r}-3)]^2},\\
     \hat{\Omega}_{3,\textrm{TD}}^{'}(\hat{r},\hat{a})&=\frac{3(\hat{a}\mp\sqrt{\hat{r}})[\pm204\hat{a}^6-528\hat{a}^5\sqrt{\hat{r}}\pm257\hat{a}^4\hat{r}+3\hat{a}^3(39+164\hat{a}^2)\sqrt{\hat{r}^3}\mp786\hat{a}^4\hat{r}^2-570\hat{a}^3\sqrt{\hat{r}^5}]}{16\sqrt{\hat{r}^9}(\hat{a}\pm\sqrt{\hat{r}^3})^4[2\hat{a}\pm\sqrt{\hat{r}}(\hat{r}-3)]^2}\\
     &+\frac{3(\hat{a}\mp\sqrt{\hat{r}})[\pm3\hat{a}^2\hat{r}^3(139\hat{a}^2+414)-3\hat{a}(126+89\hat{a}^2)\sqrt{\hat{r}^7}\mp871\hat{a}^2\hat{r}^4+3\hat{a}(279+40\hat{a}^2)\sqrt{\hat{r}^9}]}{16\sqrt{\hat{r}^9}(\hat{a}\pm\sqrt{\hat{r}^3})^4[2\hat{a}\pm\sqrt{\hat{r}}(\hat{r}-3)]^2}\\
     &+\frac{3(\hat{a}\mp\sqrt{\hat{r}})[\pm24\hat{r}^5(8\hat{a}^2-9)-456\hat{a}\sqrt{\hat{r}^{11}}\pm9\hat{r}^6(16+\hat{a}^2)+57\hat{a}\sqrt{\hat{r}^{13}}\mp16\hat{r}^7]}{16\sqrt{\hat{r}^9}(\hat{a}\pm\sqrt{\hat{r}^3})^4[2\hat{a}\pm\sqrt{\hat{r}}(\hat{r}-3)]^2},\\
     \hat{\Omega}_{4,\textrm{TD}}^{''}(\hat{r},\hat{a})&=\frac{3(\hat{a}\mp\sqrt{\hat{r}})[\mp12168\hat{a}^9+53340\hat{a}^8\sqrt{\hat{r}}\mp63642\hat{a}^7\hat{r}-\hat{a}^6(18143+42804\hat{a}^2)\sqrt{\hat{r}^3}\pm3\hat{a}^5\hat{r}^2(23263+48860\hat{a}^2)]}{128\sqrt{\hat{r}^{13}}(\hat{a}\pm\sqrt{\hat{r}^3})^5[2\hat{a}\pm\sqrt{\hat{r}}(\hat{r}-3)]^3}\\
 &+\frac{3(\hat{a}\mp\sqrt{\hat{r}})[-9\hat{a}^4(2489+8581\hat{a}^2)\sqrt{\hat{r}^5}\mp9\hat{a}^3\hat{r}^3(711+24466\hat{a}^2+6730\hat{a}^4)+3\hat{a}^4(94516+47597\hat{a}^2)\sqrt{\hat{r}^7}]}{128\sqrt{\hat{r}^{13}}(\hat{a}\pm\sqrt{\hat{r}^3})^5[2\hat{a}\pm\sqrt{\hat{r}}(\hat{r}-3)]^3}\\
 &+\frac{3(\hat{a}\mp\sqrt{\hat{r}})[\pm18\hat{a}^3\hat{r}^4(-4379+3341\hat{a}^2)-3\hat{a}^2(3753+120505\hat{a}^2+14409\hat{a}^4)\sqrt{\hat{r}^9}\pm3\hat{a}^3\hat{r}^5(81307+15062\hat{a}^2)]}{128\sqrt{\hat{r}^{13}}(\hat{a}\pm\sqrt{\hat{r}^3})^5[2\hat{a}\pm\sqrt{\hat{r}}(\hat{r}-3)]^3}\\
 &+\frac{3(\hat{a}\mp\sqrt{\hat{r}})[135\hat{a}^2(-16+1127\hat{a}^2)\sqrt{\hat{r}^{11}}\mp27\hat{a}\hat{r}^6(879+8832\hat{a}^2+593\hat{a}^4)-3\hat{a}^2(-17353+3244\hat{a}^2)\sqrt{\hat{r}^{13}}]}{128\sqrt{\hat{r}^{13}}(\hat{a}\pm\sqrt{\hat{r}^3})^5[2\hat{a}\pm\sqrt{\hat{r}}(\hat{r}-3)]^3}\\
 &+\frac{3(\hat{a}\mp\sqrt{\hat{r}})[\pm9\hat{a}\hat{r}^7(5038+9105\hat{a}^2)-3(4347+16187\hat{a}^2+879\hat{a}^4)\sqrt{\hat{r}^{15}}\mp6\hat{a}\hat{r}^8(3487+1541\hat{a}^2)]}{128\sqrt{\hat{r}^{13}}(\hat{a}\pm\sqrt{\hat{r}^3})^5[2\hat{a}\pm\sqrt{\hat{r}}(\hat{r}-3)]^3}\\
 &+\frac{3(\hat{a}\mp\sqrt{\hat{r}})[9(1601+
 1768\hat{a}^2)\sqrt{\hat{r}^{17}}\mp9\hat{a}\hat{r}^9(-206+15\hat{a}^2)-3(1961+429\hat{a}^2)\sqrt{\hat{r}^{19}}\pm75\hat{a}\hat{r}^{10}+971\sqrt{\hat{r}^{21}}]}{128\sqrt{\hat{r}^{13}}(\hat{a}\pm\sqrt{\hat{r}^3})^5[2\hat{a}\pm\sqrt{\hat{r}}(\hat{r}-3)]^3},\\
  \hat{\Omega}_{4,\textrm{TD}}^{'''}(\hat{r},\hat{a})&=\frac{3(\hat{a}\mp\sqrt{\hat{r}})[\mp3780\hat{a}^8+11784\hat{a}^7\sqrt{\hat{r}}\mp9561\hat{a}^6\hat{r}-\hat{a}^5(2497+8856\hat{a}^2)\sqrt{\hat{r}^3})]}{128\sqrt{\hat{r}^{13}}(\hat{a}\pm\sqrt{\hat{r}^3})^5[2\hat{a}\pm\sqrt{\hat{r}}(\hat{r}-3)]^2}\\
 &+\frac{3(\hat{a}\mp\sqrt{\hat{r}})[\pm3\hat{a}^4\hat{r}^2(2563+6062\hat{a}^2)-9\hat{a}^3(515+107\hat{a}^2)\sqrt{\hat{r}^5}\mp\hat{a}^4\hat{r}^3(20119+8721\hat{a}^2)]}{128\sqrt{\hat{r}^{13}}(\hat{a}\pm\sqrt{\hat{r}^3})^5[2\hat{a}\pm\sqrt{\hat{r}}(\hat{r}-3)]^2}\\
 &+\frac{3(\hat{a}\mp\sqrt{\hat{r}})[3\hat{a}^3(3963\hat{a}^2+8237)\sqrt{\hat{r}^7}\pm3\hat{a}^2\hat{r}^4(508\hat{a}^2-4653)-\hat{a}^3(4509\hat{a}^2+13312)\sqrt{\hat{r}^9}\pm3\hat{a}^2\hat{r}^5(8535+1339\hat{a}^2)]}{128\sqrt{\hat{r}^{13}}(\hat{a}\pm\sqrt{\hat{r}^3})^5[2\hat{a}\pm\sqrt{\hat{r}}(\hat{r}-3)]^2}\\
 &+\frac{3(\hat{a}\mp\sqrt{\hat{r}})[15\hat{a}(-999+40\hat{a}^2)\sqrt{\hat{r}^{11}}\mp\hat{a}^2\hat{r}^6(1215\hat{a}^2+12916)+3\hat{a}(113\hat{a}^2+7471)\sqrt{\hat{r}^{13}}\pm9\hat{r}^7(-845+213\hat{a}^2)]}{128\sqrt{\hat{r}^{13}}(\hat{a}\pm\sqrt{\hat{r}^3})^5[2\hat{a}\pm\sqrt{\hat{r}}(\hat{r}-3)]^2}\\
     &+\frac{3(\hat{a}\mp\sqrt{\hat{r}})[-\hat{a}(8743+135\hat{a}^2)\sqrt{\hat{r}^{15}}\mp3\hat{r}^8(-2234+45\hat{a}^2)+723\hat{a}\sqrt{\hat{r}^{17}}\mp1453\hat{r}^9]}{128\sqrt{\hat{r}^{13}}(\hat{a}\pm\sqrt{\hat{r}^3})^5[2\hat{a}\pm\sqrt{\hat{r}}(\hat{r}-3)]^2},\\
      \hat{\Omega}_{3,\textrm{TD}}^{''}(\hat{r},\hat{a})&=\frac{3(\hat{a}\mp\sqrt{\hat{r}})[\pm204\hat{a}^6-624\hat{a}^5\sqrt{\hat{r}}\pm401\hat{a}^4\hat{r}+3\hat{a}^3(103+164\hat{a}^2)\sqrt{\hat{r}^3}\mp18\hat{a}^2\hat{r}^2(16+57\hat{a}^2)-378\hat{a}^3\sqrt{\hat{r}^5}]}{16\sqrt{\hat{r}^9}(\hat{a}\pm\sqrt{\hat{r}^3})^4[2\hat{a}\pm\sqrt{\hat{r}}(\hat{r}-3)]^2}\\
     &+\frac{3(\hat{a}\mp\sqrt{\hat{r}})[\pm3\hat{a}^2\hat{r}^3(139\hat{a}^2+622)-9\hat{a}(106+51\hat{a}^2)\sqrt{\hat{r}^7}\mp967\hat{a}^2\hat{r}^4+3\hat{a}(503+40\hat{a}^2)\sqrt{\hat{r}^9}]}{16\sqrt{\hat{r}^9}(\hat{a}\pm\sqrt{\hat{r}^3})^4[2\hat{a}\pm\sqrt{\hat{r}}(\hat{r}-3)]^2}\\
     &+\frac{3(\hat{a}\mp\sqrt{\hat{r}})[\pm72\hat{r}^5(2\hat{a}^2-7)-648\hat{a}\sqrt{\hat{r}^{11}}\pm3\hat{r}^6(128+3\hat{a}^2)+57\hat{a}\sqrt{\hat{r}^{13}}\mp64\hat{r}^7]}{16\sqrt{\hat{r}^9}(\hat{a}\pm\sqrt{\hat{r}^3})^4[2\hat{a}\pm\sqrt{\hat{r}}(\hat{r}-3)]^2}.\\
\end{align*}
\end{widetext} \newpage

\begin{widetext}
\begin{align*}
     \hat{\Omega}_{4,\textrm{TD}}^{''''}(\hat{r},\hat{a})&=\frac{3(\hat{a}\mp\sqrt{\hat{r}})[\pm768\hat{a}^8-3840\hat{a}^7\sqrt{\hat{r}}\pm6144\hat{a}^6\hat{r}+3072\hat{a}^5(\hat{a}-1)(\hat{a}+1)\sqrt{\hat{r}^3})]}{128\sqrt{\hat{r}^{17}}(\hat{a}\pm\sqrt{\hat{r}^3})^5[2\hat{a}\pm\sqrt{\hat{r}}(\hat{r}-3)]^2}\\
 &+\frac{3(\hat{a}\mp\sqrt{\hat{r}})[\mp12\hat{a}^6\hat{r}^2(1088+315\hat{a}^2)+24\hat{a}^5(736+475\hat{a}^2)\sqrt{\hat{r}^5}\mp3\hat{a}^4\hat{r}^3(2560+1331\hat{a}^2)]}{128\sqrt{\hat{r}^{17}}(\hat{a}\pm\sqrt{\hat{r}^3})^5[2\hat{a}\pm\sqrt{\hat{r}}(\hat{r}-3)]^2}\\
 &+\frac{3(\hat{a}\mp\sqrt{\hat{r}})[-\hat{a}^5(8856\hat{a}^2+17665)\sqrt{\hat{r}^7}\pm3\hat{a}^4\hat{r}^4(5614\hat{a}^2+7043)+9\hat{a}^3(533\hat{a}^2-899)\sqrt{\hat{r}^9}\mp\hat{a}^4\hat{r}^5(24343+8721\hat{a}^2)]}{128\sqrt{\hat{r}^{17}}(\hat{a}\pm\sqrt{\hat{r}^3})^5[2\hat{a}\pm\sqrt{\hat{r}}(\hat{r}-3)]^2}\\
 &+\frac{3(\hat{a}\mp\sqrt{\hat{r}})[3\hat{a}^3(6509+3387\hat{a}^2)\sqrt{\hat{r}^{11}}\pm3\hat{a}^2\hat{r}^6(1468\hat{a}^2-2989)-\hat{a}^3(4509\hat{a}^2+7744)\sqrt{\hat{r}^{13}}]}{128\sqrt{\hat{r}^{17}}(\hat{a}\pm\sqrt{\hat{r}^3})^5[2\hat{a}\pm\sqrt{\hat{r}}(\hat{r}-3)]^2}\\
     &+\frac{3(\hat{a}\mp\sqrt{\hat{r}})[\pm3\hat{a}^2\hat{r}^7(4631+1019\hat{a}^2)+3\hat{a}(136\hat{a}^2-3331)\sqrt{\hat{r}^{15}}\mp\hat{a}^2\hat{r}^8(7348+1215\hat{a}^2)+3\hat{a}(5615+49\hat{a}^2)\sqrt{\hat{r}^{17}}]}{128\sqrt{\hat{r}^{17}}(\hat{a}\pm\sqrt{\hat{r}^3})^5[2\hat{a}\pm\sqrt{\hat{r}}(\hat{r}-3)]^2}\\
     &+\frac{3(\hat{a}\mp\sqrt{\hat{r}})[\pm3\hat{r}^9(383\hat{a}^2-2151)-\hat{a}(6823+135\hat{a}^2)\sqrt{\hat{r}^{19}}\mp9\hat{r}^{10}(15\hat{a}^2-638)+531\hat{a}\sqrt{\hat{r}^{21}}\mp1261\hat{r}^{11}]}{128\sqrt{\hat{r}^{17}}(\hat{a}\pm\sqrt{\hat{r}^3})^5[2\hat{a}\pm\sqrt{\hat{r}}(\hat{r}-3)]^2},\\
      \hat{\Omega}_{3,\textrm{TD}}^{'''}(\hat{r},\hat{a})&=\frac{3(\hat{a}\mp\sqrt{\hat{r}})[96\hat{a}^7\mp480\hat{a}^6\sqrt{\hat{r}}+768\hat{a}^5\hat{r}\pm96\hat{a}^4(3\hat{a}^2-4)\sqrt{\hat{r}^3}-1152\hat{a}^5\hat{r}^2\pm12\hat{a}^4(120+19\hat{a}^2)\sqrt{\hat{r}^5}]}{16\hat{r}^7(\hat{a}\pm\sqrt{\hat{r}^3})^4[2\hat{a}\pm\sqrt{\hat{r}}(\hat{r}-3)]^2}\\
     &+\frac{3(\hat{a}\mp\sqrt{\hat{r}})[-72\hat{a}^3\hat{r}^3(5\hat{a}^2+8)\mp415\hat{a}^4\sqrt{\hat{r}^7}+3\hat{a}^3\hat{r}^4(279+172\hat{a}^2)\mp6\hat{a}^2(48+155\hat{a}^2)\sqrt{\hat{r}^9}-450\hat{a}^3\hat{r}^5]}{16\hat{r}^7(\hat{a}\pm\sqrt{\hat{r}^3})^4[2\hat{a}\pm\sqrt{\hat{r}}(\hat{r}-3)]^2}\\
     &+\frac{3(\hat{a}\mp\sqrt{\hat{r}})[\pm3\hat{a}^2(542+131\hat{a}^2)\sqrt{\hat{r}^{11}}-3\hat{a}\hat{r}^6(254+137\hat{a}^2)\mp847\hat{a}^2\sqrt{\hat{r}^{13}}+3\hat{a}\hat{r}^7(455+32\hat{a}^2)]}{16\hat{r}^7(\hat{a}\pm\sqrt{\hat{r}^3})^4[2\hat{a}\pm\sqrt{\hat{r}}(\hat{r}-3)]^2}\\
     &+\frac{3(\hat{a}\mp\sqrt{\hat{r}})[\pm72(2\hat{a}^2-7)\sqrt{\hat{r}^{15}}-576\hat{a}\hat{r}^8\pm3(112+3\hat{a}^2)\sqrt{\hat{r}^{17}}+33\hat{a}\hat{r}^9\mp40\sqrt{\hat{r}^{19}}]}{16\hat{r}^7(\hat{a}\pm\sqrt{\hat{r}^3})^4[2\hat{a}\pm\sqrt{\hat{r}}(\hat{r}-3)]^2}.
\end{align*}
\end{widetext}

\bibliographystyle{unsrt}
\bibliography{refs1}

\end{document}